\documentclass[aps,prx,superscriptaddress,showpacs,twocolumn]{revtex4-2}
\usepackage[left=1.6cm,right=1.6cm,bottom=1.7cm,top=1.7cm,ignoreall]{geometry}
\usepackage[utf8]{inputenc}
\usepackage{graphicx}
\usepackage{amsmath}
\usepackage{amssymb}
\usepackage{xspace}
\usepackage{bm}
\usepackage{color}
\usepackage[squaren,Gray]{SIunits}
\usepackage[hyperindex=true]{hyperref}
\hypersetup{linktocpage,colorlinks=true,citecolor=blue,linkcolor=blue}
\usepackage{empheq}
\usepackage{braket}
\usepackage{physics}

\begin{document}

\title{Tunable Generation of Spatial Entanglement in Nonlinear Waveguide Arrays}	

\author{A.\ Raymond}
\affiliation{Université Paris Cité, CNRS, Laboratoire Matériaux et Phénomènes Quantiques, 75013 Paris, France}

\author{A.\ Zecchetto}
\affiliation{Université Paris Cité, CNRS, Laboratoire Matériaux et Phénomènes Quantiques, 75013 Paris, France}

\author{J.~Palomo}
\affiliation{ENS-PSL, Université PSL, CNRS, Sorbonne Université, Laboratoire de Physique de l’Ecole Normale Supérieure, 75005 Paris, France}

\author{M.~Morassi}
\affiliation{Université Paris-Saclay, CNRS, Centre de Nanosciences et de Nanotechnologies, 91120 Palaiseau, France}

\author{A.~Lemaître}
\affiliation{Université Paris-Saclay, CNRS, Centre de Nanosciences et de Nanotechnologies, 91120 Palaiseau, France}

\author{F.~Raineri}
\affiliation{Université Côte d’Azur, Institut de Physique de Nice, CNRS-UMR 7010, 06200 Nice, France}
\affiliation{Université Paris-Saclay, CNRS, Centre de Nanosciences et de Nanotechnologies, 91120 Palaiseau, France}

\author{M.I.~Amanti}
\affiliation{Université Paris Cité, CNRS, Laboratoire Matériaux et Phénomènes Quantiques, 75013 Paris, France}

\author{S.~Ducci}
\affiliation{Université Paris Cité, CNRS, Laboratoire Matériaux et Phénomènes Quantiques, 75013 Paris, France}

\author{F.~Baboux}
\thanks{florent.baboux@u-paris.fr}
\affiliation{Université Paris Cité, CNRS, Laboratoire Matériaux et Phénomènes Quantiques, 75013 Paris, France}

\makeatletter
\def\Dated@name{} 
\makeatother

\date{}

\begin{abstract}

Harnessing high-dimensional entangled states of light presents a frontier for advancing quantum information technologies, from fundamental tests of quantum mechanics to enhanced computation and communication protocols. In this context, the spatial degree of freedom stands out as particularly suited for on-chip integration. But while traditional demonstrations produce and manipulate path-entangled states sequentially with discrete optical elements, continuously-coupled nonlinear waveguide systems offer a promising alternative where photons can be generated and interfere along the entire propagation length, unveiling novel capabilities within a reduced footprint. Here we exploit this concept to implement a compact and reconfigurable source of path-entangled photon pairs based on parametric down-conversion in semiconductor nonlinear waveguides arrays. We use a double-pump configuration to engineer the output quantum state and implement various types of spatial correlations, exploiting a quantum interference effect between the biphoton state generated in each pumped waveguide. This demonstration, at room temperature and telecom wavelength, illustrates the potential of continuously-coupled systems as a promising alternative to discrete multi-component quantum circuits for leveraging the high-dimensional spatial degree of freedom of photons.

\end{abstract}

\maketitle

Nonclassical states of light constitute crucial resources for quantum information technologies due to their ability to transmit easily, resist decoherence, and offer various ways to encode information \cite{Walmsley15}. Recently, considerable attention has been devoted to entanglement in high-dimensional degrees of freedom of photons \cite{Erhard20} to push fundamental tests of quantum mechanics \cite{Dada11,Yu24}, boost the efficiency and security of quantum communication~\cite{Ding17,Cozzolino19}, and make quantum computing more flexible \cite{Reimer19,Paesani21}.

Among the various candidates, the spatial degree of freedom emerges as especially fitted for on-chip integration \cite{Solntsev17,Wang20}. Rapid progress has thus been made in recent years to develop complex integrated circuits achieving on-chip quantum interference \cite{Politi08}, entanglement and various logical operations on path-encoded states \cite{Carolan15,Wang18}, in the spirit of the gate-based model of  quantum information processing \cite{Knill01}. 
These powerful demonstrations typically implement a discrete and sequential manipulation of path-entangled states using beamplitters and phase shifters, with most of the footprint devoted to routing waveguides. 

Another route to spatial entanglement is however possible, based on continuously-coupled waveguides \cite{Perets08,Bromberg09} where photons can interfere across the entire propagation length rather than solely at individual beamsplitters.
This approach implements continuous-time random quantum walks \cite{Farhi98} and can be described by a lattice Hamiltonian \cite{Perets08}, establishing a natural connection with a variety of situations encountered in condensed matter physics. Arrays of coupled waveguides have recently facilitated investigations into a large spectrum of phenomena spanning quantum walks of correlated photons \cite{Peruzzo10,Jiao21,Hoch22}, Anderson localization of photon pairs \cite{DiGiuseppe13}, quantum logic operations \cite{Lahini18,Chapman24}, topological effects \cite{Tambasco18,Klauck21} or non-Hermitian physics \cite{Gao24}. These remarkable achievements have relied on external sources to generate quantum states of light, which were then fed into a passive circuitry.

Yet, an additional layer of complexity can introduce even larger possibilities, by adding to the continuous nature of the interference process the ability to continuously generate photons all along the device. 
This can be realized in a waveguide array made out of a nonlinear material, by injecting a classical pump beam generating stochastically photon pairs by spontaneous parametric down-conversion (SPDC) or four-wave mixing \cite{Solntsev12,Grafe12,Kruse13,Solntsev14,BlancoRedondo18,Bergamasco19,Doyle22}. 
This approach offers novel configurations that have no equivalent in bulk optics nor discrete photonic circuits due to the intricate combination of the generation and manipulation steps of photonic quantum states \cite{Luo19,Belsley20,Barral20,Barral21,He24}.

\begin{figure*}[t]
\centering
\includegraphics[width=0.96\textwidth]{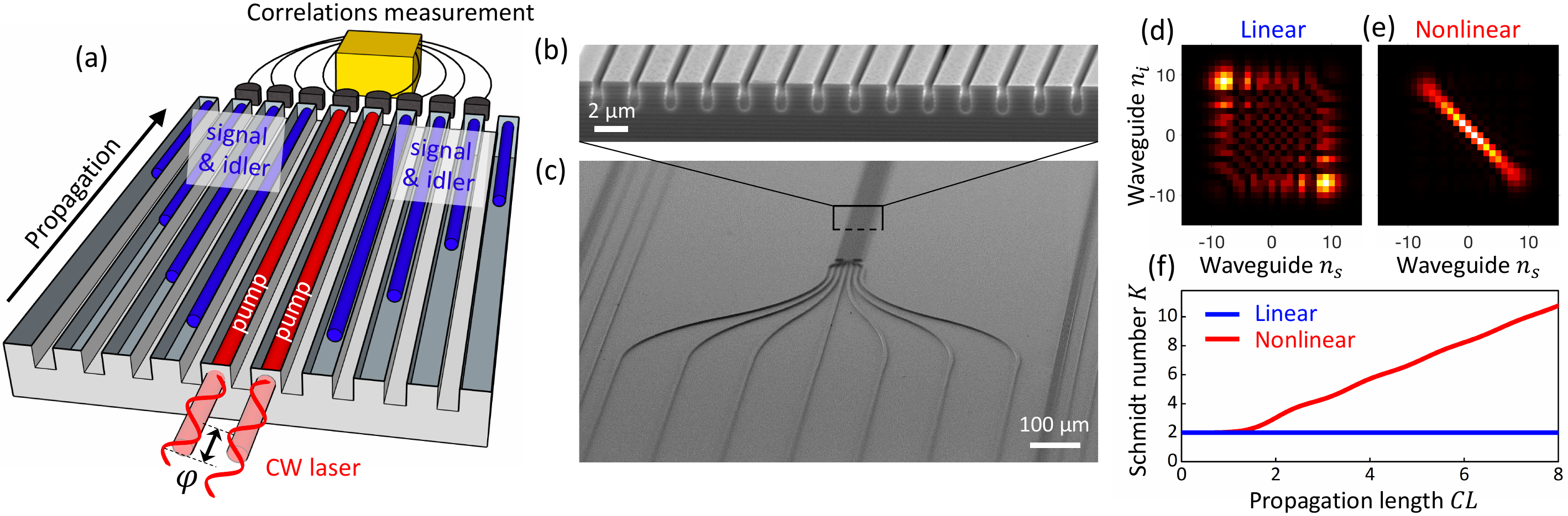}
\caption{
(a) Principle of the spatial entanglement engineering in a quadratic nonlinear waveguide array. Here, two continuous-wave pump laser beams (sketched in red) with a phase offset $\varphi$ between them generate photon pairs that undergo quantum walks (blue), resulting in path entanglement over the various waveguides of the array. 
(b) SEM image of a fabricated AlGaAs sample, showing a cut of the waveguide array in the central part.
(c) Wider-field SEM image of the sample.
(d) Simulated correlation matrix for a linear waveguide array injected with a path-entangled state $ (\ket{00} + \ket{11} )/\sqrt{2}$, and (e) for a nonlinear array when pumping waveguides $n=0$ and $1$ in-phase. (f)~Corresponding calculated Schmidt number as a function of the (normalized) propagation length.}
\label{Fig1}
\end{figure*}

In this Letter, we exploit this concept in AlGaAs-based nonlinear waveguide arrays to implement a compact and versatile source of path-entangled states of light, operating at room temperature and telecom wavelength. We use a double-pump configuration to engineer the output quantum state and implement different types of spatial correlations, violating a non-classicality criterion by several tens of standard deviations. This control relies on a quantum interference effect between the biphoton wavefunction generated in each pumped waveguide, fully accounted for by analytical calculations and numerical simulations, and paving the way to quantum information tasks leveraging the high-dimensional spatial degree of freedom in continuously-coupled photonic systems.

The working principle of our semiconductor AlGaAs nonlinear waveguide array is sketched in Fig.~\ref{Fig1}a. A classical pump beam (wavelength $\simeq 775$ nm, sketched in red) injected into one or several waveguides generates photon pairs at telecom wavelength ($\simeq 1550$ nm, shown in blue) by SPDC, thanks to the strong $\chi^{(2)}$ nonlinearity of the material. 
These photon pairs can continuously tunnel from one waveguide to the other during their propagation, implementing random quantum walks. Compared to quantum walks in passive circuits \cite{Peruzzo10,Jiao21,Hoch22}, the walkers are here generated directly inside the device and the generation can take place at any position along the propagation axis \cite{Solntsev12}.
Besides a gain of integration, this configuration allows for a progressive buildup of spatial entanglement, due to the interference between quantum walks initiated at all possible longitudinal positions, as we shall detail in the following.
The device is designed so that the transverse coupling of the pump beam is negligible, so that the pump remains confined in the initially pumped waveguide(s) during propagation \cite{SM}.

The precise articulation between quantum states produced by linear waveguide arrays (i.e. injected with photons produced externally) and nonlinear arrays (where photon pairs are generated continuously by internal SPDC) can be determined exactly through analytical calculations.
Let $\ket{\Psi}  \!=\!\sum_{n_s,n_i} \Psi (n_{s},n_{i}) \ket{n_s,n_i} $ denote the biphoton state generated by a nonlinear array, where $\Psi (n_{s},n_{i}) $ governs the probability amplitude to detect one photon in waveguide number $n_s$ and the other photon in waveguide $n_i$.
If the nonlinear array is pumped by a monochromatic pump field tuned to the single-waveguide phase matching condition, generating photon pairs by frequency-degenerate SPDC, one can show \cite{SM} that 
\begin{equation}
\Psi (n_{s},n_{i}) = \gamma \sum_n A_n \!\int_{0}^{L} \!\! \phi_n(n_{s},n_{i},z)  \,  dz 
\label{Active}
\end{equation}
where $\gamma$ accounts for the efficiency of the SPDC process, $z$ is the longitudinal position, $L$ is the array length, and $A_n$ is the pump field amplitude in waveguide $n$. Interestingly, $\phi_n(n_{s},n_{i},z) $ corresponds to
the quantum state that would be produced by a \textit{linear} array with two photons initially injected in waveguide $n$, after a propagation distance $z$~\cite{SM}. The latter state reads
\begin{equation}
	\phi_n(n_{s},n_{i},z) = i^{n_{s} + n_{i}-2n}\ J_{n_{s}-n}(2Cz)\ J_{n_{i}-n}(2Cz)
\end{equation}
where  $J_{n}$ is the first-order Bessel function of index $n$ and $C$ is the coupling constant between waveguides \cite{Bromberg09}.

Thus, the output state of a nonlinear waveguide array can be seen as resulting from the interference (weighted by the pump spatial distribution $A_n$) of the biphoton state $\Psi_n (n_{s},n_{i}) =  \gamma \int_{0}^{L} \! \phi_n(n_{s},n_{i},z)  \,  dz$ generated in each pumped waveguide. In turn, since the SPDC photons can be generated at all possible longitudinal positions, the state $\Psi_n (n_{s},n_{i})$ results from the interference of states $\phi_n(n_{s},n_{i},z)$ corresponding to propagation in linear arrays of continuously varying lengths between $0$ and $L$.
The spatial correlations in real space are described by a (symmetric) matrix $\Gamma_{n_s,n_i}  = |\Psi(n_s,n_i)|^2$. Note that $\Gamma_{n_s,n_i} $ is not normalized to unity, it takes into account the SPDC efficiency and hence is directly proportional to coincidence counts as will be measured experimentally in the following.

To illustrate this interplay between quantum walks and nonlinearity, let us compare  \textbf{(i)} the output state of a linear waveguide array injected with a path-entangled state $ (\ket{00} + \ket{11} )/\sqrt{2}$ (corresponding to two indistinguishable photons coupled in either the central waveguide $n\!=\!0$ or the neighboring waveguide $n\!\!=\!\!1$), and \textbf{(ii)} the state generated by a nonlinear array when pumping these two waveguides with equal amplitude ($A_0\!=\!A_1$).
The corresponding intensity correlation maps are plotted in Fig.~\ref{Fig1}d and e, respectively, for $CL=5$.
In the linear case (i), photons are distributed over a square-like pattern, with dominant antidiagonal lobes resulting in a spatially anticorrelated state.
These lobes, located near $\pm 2 CL$, correspond to a ballistic-like propagation of both photons in opposite directions \cite{Bromberg09}.
As the propagation length increases, this overall pattern expands linearly, with the biphoton wavefunction spreading over more waveguides. However the entanglement level of the state, quantified by the Schmidt number $K$ (Fig.~\ref{Fig1}f, blue), remains fixed to that of the injected state ($K\!=\!2$) since propagation in a linear optical circuit cannot increase entanglement \cite{Sperling11}.

Now turning to the nonlinear case (ii), the output state (Fig.~\ref{Fig1}e) results from the interference between linear-array states (i) obtained for continuously varying propagation lengths (Eq.~\eqref{Active}). This gives rise to a sharp interference pattern, where essentially only antidiagonal points survive.
As the propagation length increases, the Schmidt number builds up monotonically (Fig.~\ref{Fig1}f, red). Obtaining such high-dimensional entangled states from a linear array would necessitate the delicate preparation of a quantum state of same $K$ at the input. 
In nonlinear arrays by contrast, the constructive interplay between quantum walks and SPDC generation provides a compact source of spatial entanglement, which can be entirely controlled from simple classical resources, namely from the spatial distribution of the input pump laser. In this work, we experimentally demonstrate this concept using a double-pump configuration.

\color{black}

The investigated sample consists in an array of 31 coupled AlGaAs waveguides of length $L=2$ mm (Fig.~\ref{Fig1}b). The waveguides are 1.7 $\mu$m-wide and separated by 450~nm gaps, resulting in a coupling constant $C_{\rm TE}=2.7$ mm$^{-1}$ in TE polarization and $C_{\rm TM}=2.4$ mm$^{-1}$ in TM polarization.
The 7 central waveguides of the array are connected to fan-in injection waveguides and fan-out collection waveguides to facilitate their individual optical addressing (Fig.~\ref{Fig1}c). These injection and collection waveguides have a larger width (6 $\mu$m) to shift their nonlinear resonance wavelength so that the injected pump laser generates photon pairs only in the central array region,  through a modal phase-matching scheme \cite{SM,Helmy11,Horn12,Boitier14,Baboux23}.

The experiments are carried out by injecting a TE-polarized continuous-wave pump laser with a microscope objective into one or several waveguides (depending on the targeted quantum state), to generate orthogonally polarized signal and idler photons by type-2 SPDC. These SPDC photons undergo random quantum walks and are collected at the output of the fan-out waveguides using a commercial lensed fiber array connected to superconducting nanowire single-photon detectors (SNSPD). 
Correlation events, measured by a time tagger, yield the spatial (path) correlation matrix $\Gamma_{n_s,n_i}$.

\begin{figure*}[t]
\centering
\includegraphics[width=0.96\textwidth]{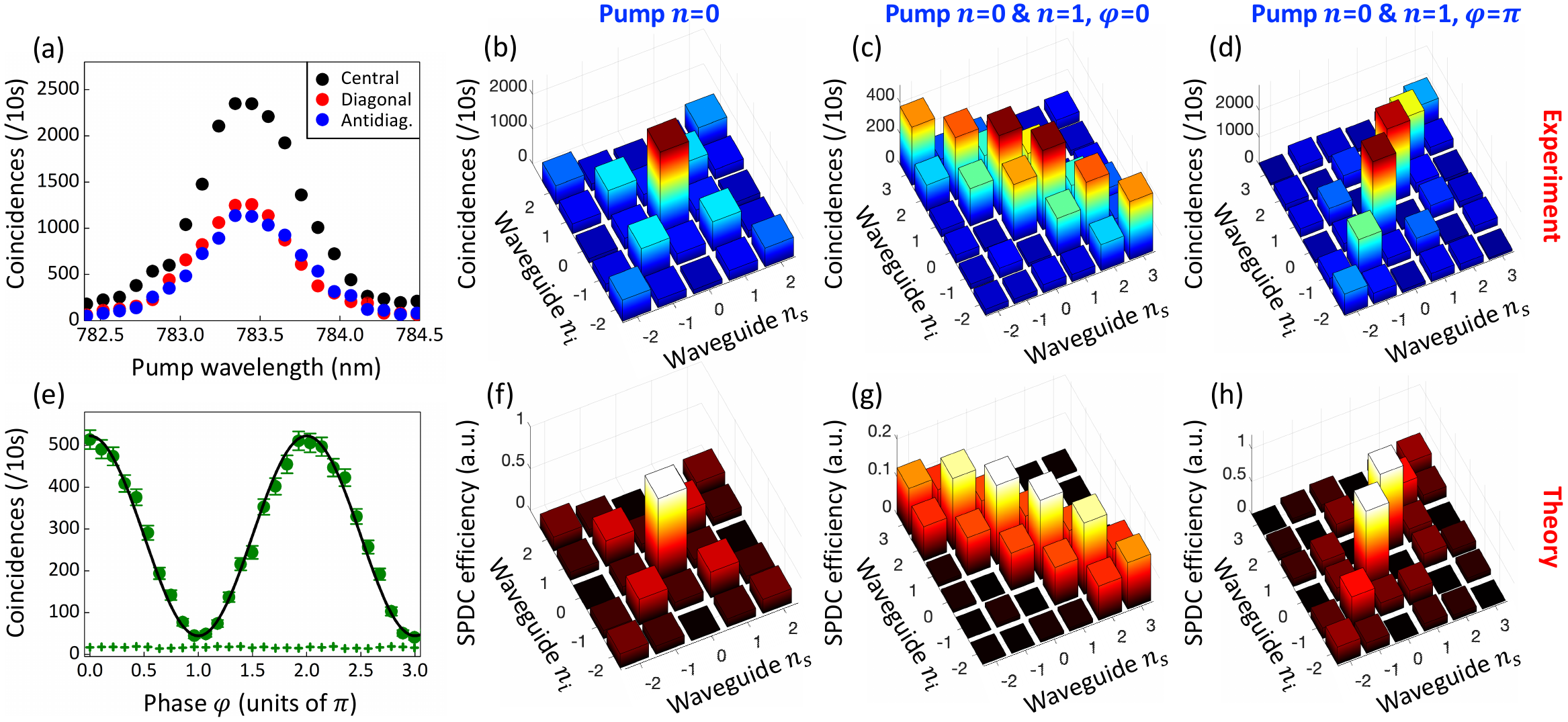}
\caption{
(a) Biphoton correlations measured in the central waveguide ($\Gamma_{0,0}$, black), first diagonal ($\Gamma_{1,1}+\Gamma_{-1,-1}$, red) and first antidiagonal ($\Gamma_{1,-1}+\Gamma_{-1,1}$, blue) points of the correlation matrix when pumping the $n=0$ waveguide (raw coincidence counts).
(b-c-d) Measured correlation matrices when pumping (b) the central waveguide $n=0$, (c) guides $n=0$ and $n=1$ in phase and (d) guides $n=0$  and $n=1$  out of phase. 
(e) Cross-correlation term $\Gamma_{0,1}$ measured for varying phase offset $\varphi$ between the two pumped waveguides ($n=0$  and $1$); circles show raw coincidence counts with Poissonian error bars, crosses show noise counts, and the black line is a sinusoidal fit.
(f-g-h) Numerically simulated correlation matrices corresponding to the experimental cases (b-c-d).
}
\label{Fig2}
\end{figure*}

The pump laser is first injected into the central waveguide ($n=0$) of the array, and coincidences are monitored as a function of the laser wavelength for a coupled pump power of 1 mW. Here and in the subsequent measurements, a bandpass filter 
is used before photon detection to reduce the emission bandwidth to 16 nm and effectively suppress any possible coupling between the spatial and spectral properties of the biphoton states~\cite{SM}.
As shown in Fig.~\ref{Fig2}a, the coincidences in the central waveguide ($\Gamma_{0,0}$, black points) show a resonance centered at $\lambda_{p,0}=783.45$ nm, which corresponds (up to 0.05 nm) to the resonance wavelength measured for a single (uncoupled) waveguide.
The diagonal coincidences ($\Gamma_{1,1}+\Gamma_{-1,-1}$, red points) and antidiagonal coincidences ($\Gamma_{1,-1}+\Gamma_{-1,1}$, blue points) display a similar resonant behavior.

We next set the pump wavelength to $\lambda_{p,0}$ and proceed to the measurement of the full correlation matrix within the Hilbert space spanned by the 5 central waveguides. The result is shown in Fig.~\ref{Fig2}b and compared to numerical simulations in Fig.~\ref{Fig2}f.
As anticipated by the results of Fig.~\ref{Fig2}a, we observe strong and selective correlations along the diagonal and the antidiagonal axes. That is, photons have an enhanced probability to exit the device either through the same waveguide ($n_s=n_i$, spatial bunching) or through opposite waveguides ($n_s=-n_i$, spatial antibunching). 
The non-classicality of the measured correlations can be quantified by using the criteria derived by Bromberg et al.~\cite{Bromberg09,Peruzzo10}. 
For classical light, off-diagonal correlations  ($\Gamma_{n_s,n_i}$ with $n_s \neq n_i$) are related to diagonal correlations by the inequality $\Gamma_{n_s,n_i} > \frac{2}{3}\sqrt{\Gamma_{n_s,n_s}\Gamma_{n_i,n_i}}$.
For the state of Fig.~\ref{Fig2}b, this inequality is violated (up to 20 standard deviations) by several points of the measured correlation matrix~\cite{SM}, indicating spatial correlations that cannot be generated by classical means. 
The overlap between the experimental ($\Gamma_{n_s,n_i} ^{\rm exp} $) and theoretical ($\Gamma_{n_s,n_i} ^{\rm th} $) correlations matrices can be quantified using the similarity parameter~\cite{Peruzzo10,Jiao21} $S= \big( \sum \sqrt{ \Gamma_{n_s,n_i} ^{\rm exp}  \Gamma_{n_s,n_i} ^{\rm th}}  \big)^2 / \left(  \sum  \Gamma_{n_s,n_i} ^{\rm exp} \sum  \Gamma_{n_s,n_i} ^{\rm th}   \right)$, 
where the summations run over all indices $n_s$ and $n_i$.
We obtain $S=98.4 \pm 1.1 \%$ (without adjustable parameter), illustrating the high fidelity of the source. 
We note that the simulation of Fig.~\ref{Fig2}f is exact, i.e.\ it takes into account the finite number of waveguides, the finite bandwidth of the photon pairs and the slight polarization and frequency-dependence of the coupling constant; it is however very close to the result of Eq.~\eqref{Active} (mutual similarity $98.8 \%$), validating the relevance of our analytical model. 
The measured correlation map is robust to the pump power; as the chip length increases, correlations spread across farther waveguides but the overall pattern remains qualitatively unchanged (after a short stabilization regime) \cite{SM}.

We will now show how this map of spatial correlations can be flexibly engineered by tailoring the pumping configuration. For this we coherently pump two neighboring waveguides ($n=0$ and $1$) by splitting the CW input laser into an inteferometer placed before the injection objective, allowing to inject two laser beams of equal intensity (0.5 mW) and controlled phase $\varphi$ between them (see Fig.~\ref{Fig1}a). 
This double-pump configuration is first investigated by monitoring the crossed coincidences $\Gamma_{0,1}$ between the two pumped waveguides as a function of the phase $\varphi$, as shown in Fig.~\ref{Fig2}e (green circles). An oscillating behavior is observed, well reproduced by a sinusoidal modulation (black line) with a net (raw) visibility of 90\% (85\%).
This behavior is in good qualitative agreement with our analytical calculations which predict (in the limit of an infinite lattice and perfect spectral degeneracy):
\begin{equation}
\Gamma_{0,1} \propto (1+\cos \varphi) \, ,
\label{Gamma01}
\end{equation}
corresponding to an alternation of (full) destructive and constructive interference of the biphoton wavefunction at this point of the correlation matrix~\cite{SM}.

We now set $\varphi=0$ and measure the spatial correlation matrix within the 6-waveguide basis centered on the pumped waveguides. The result is shown in Fig.~\ref{Fig2}c and compared to numerical simulations in Fig.~\ref{Fig2}g (similarity $93.1 \pm 1.8 \%$).
We observe this time strong correlations along the antidiagonal, showing that spatial antibunching is selectively enhanced by this pumping configuration. 
We also notice, compared to the single-pump case (Fig.~\ref{Fig2}b), a wider propagation of the biphoton correlations, which decrease only slowly over the 6 measured waveguides.

We then set $\varphi=\pi$, corresponding to the $n=0$ and $n=1$ waveguides pumped in phase opposition. The measured correlation matrix is shown in Fig.~\ref{Fig2}d along with simulation results in \ref{Fig2}h (similarity $96.9 \pm 1.0 \%$). In this case, we observe that spatial bunching is selectively enhanced, while the spreading of correlations is comparable to the single-pump configuration. However, the total SPDC signal (integrated over all measured points of the correlation matrix) is stronger (by a factor $\simeq$ 2.8) for $\varphi=\pi$ than for $\varphi=0$, despite the fact that the input pump power is the same. As the transverse coupling of the pump beam is negligible, the two pump beams do not interfere classically and this interference effect is purely quantum in nature: the biphoton states generated in the two pumped waveguides interfere constructively when $\varphi=\pi$ and destructively when $\varphi=0$, and it can be shown analytically within realistic assumptions~\cite{SM} that the total SPDC intensity is modulated as 
\begin{equation}
I_{\rm SPDC}=\sum_{n_s,n_i}  \Gamma_{n_s,n_i} \propto (2-\cos \varphi)
\label{GammaTot}
\end{equation}
when the phase $\varphi$ between the two pumped waveguides is varied.

This intriguing behavior, with a constructive interference occurring for out-of-phase pumping, can be understood intuitively in the following manner. In the single-pump configuration, correlations are mainly located on the diagonal and antidiagonal (Figs.~\ref{Fig2}b,f).
According to coupled-mode theory, a single photon acquires a phase of $\pi/2$ when it tunnels to a neighboring waveguide; hence, starting from its generation in the central waveguide $n=0$, the biphoton state acquires a phase $\pi$ when it moves one step along the diagonal or the antidiagonal, as this requires two single-photon jumps (either on the same or on opposite sides). 
The phase of the wavefunction thus changes by $\pi$ between each two successive points along the diagonal and antidiagonal.
Now, as shown by Eq.~\eqref{Active}, when pumping two neighboring waveguides $n=0$ and $1$, an interference occurs between the biphoton states $\ket{\Psi}_{n=0}$ and $\ket{\Psi}_{n=1}$ generated in these two waveguides.
When the two pumps are in phase (resp.\ out of phase), this interference is destructive (resp.\ constructive) on the diagonal points (which are aligned for the two states) because of the $\pi$ phase alternation structure described above, in good agreement with the experiments and simulations. This interference is only partial since the two biphoton states have different amplitudes on the interfering points; however it plays a determinant role in the total SPDC efficiency since an important part of the signal is concentrated in each pumped waveguide, providing an intuitive argument for the total SPDC intensity being maximum when $\varphi=\pi$, in agreement with Eq.~\eqref{GammaTot}.
The situation is reversed in the case of antidiagonal points. However, on the antidiagonal passing through $(0,1)$ the two biphoton states interfere with the exact same amplitude, leading to a perfect interference and thus a high spatial spreading of the correlations when $\varphi=0$, in good agreement with Figs.~\ref{Fig2}c and \ref{Fig2}g. 
This demonstrated quantum interference effect provides the building block for engineering path-entangled states in continuously-coupled
nonlinear waveguides.

In summary, we have demonstrated a compact and versatile source of spatial entanglement based on quantum walks in semiconductor AlGaAs nonlinear waveguide arrays. Tailoring the spatial profile of the pump field allows to reconfigure the output quantum state and implement various types of spatial correlations, well accounted for by analytical calculations and numerical simulations.
Building upon these results, a wider zoology of quantum states could be produced by pumping more waveguides, with controlled intensity and phase relation between them, via on-chip phase-shifters exploiting the strong electro-optic effect of \mbox{AlGaAs} \cite{Wang14}.
Pushing further the integration, the possibility to integrate the pump laser directly within the nonlinear medium \cite{Baboux23} constitutes a distinctive asset of the AlGaAs platform compared to dielectric and silicon-based nonlinear photonic circuits where path-entangled states have been studied previously \cite{Solntsev17}.

In addition, the strong second-order nonlinearity of AlGaAs could be leveraged to push nonlinear waveguide arrays into the squeezing regime \cite{Brodutch18,Yan22}, enabling the generation of highly multimode squeezed states in a flexible manner \cite{Barral20,Barral21}.
Finally, other lattice geometries can be implemented by modifying the propagation and coupling constants of the waveguides either at the fabrication step (by tuning the widths and distances between waveguides) or in-situ by utilizing the electro-optic effect of AlGaAs. This would enrich the possibilities of quantum state engineering to implement e.g.~biphoton W-states \cite{Grafe14} or discrete fractional Fourier transforms \cite{Weimann16} of entangled states. The realization of quasi-periodic geometries (e.g.\ Fibonacci and Aubry-André) would allow investigating topological effects in the quantum regime \cite{Tambasco18} or the Anderson localization of multi-particle states with peculiar cascaded or reentrant behaviors \cite{Goblot20,Roy21}, making the demonstrated platform appealing to simulate in a controlled environment physical problems otherwise difficult to access in condensed-matter systems.

\vspace{-0.3cm}
\section*{Acknowledgments}
\vspace*{-0.1cm}

We thank P.~Milman, L.~Guidoni, Y.~Bromberg and I.~Carusotto for fruitful discussions and P.~Filloux for technical support. We acknowledge support from the Ville de Paris Emergence program (\textsc{LATTICE} project), Region Ile de France in the framewok of the DIM QuanTiP (\textsc{Q-LAT} project), IdEx Université Paris Cité (ANR-18-IDEX-0001), the French \textsc{RENATECH} network and the "Investissement d'Avenir" program of the French Government (ANR‐22‐CMAS-0001, QuanTEdu-France).

\vspace*{-0.2cm}

\end{document}